\documentclass[11pt]{article}

\usepackage[latin1]{inputenc}
\usepackage[T1]{fontenc}
\usepackage[english]{babel}
\usepackage{amsfonts}
\usepackage{amssymb}
\usepackage{amsmath}
\usepackage{amsthm}
\usepackage{graphicx}

\newtheorem{proposition}{Proposition}
\newtheorem{lemma}{Lemma}
\newtheorem{theorem}{Theorem}

\newtheorem{remark}{Remark}
\newcommand{\meanv}[1]{\left\langle#1\right\rangle}

\newcommand{\OO}[1]{O \left(\frac{1}{#1}\right)}

\begin{document}

\title{A Certain Class of Curie-Weiss Models}


\author{Giuseppe Genovese \footnote{Dipartimento di Fisica,
    Sapienza Universit\`a di Roma} \ and Adriano Barra$^*$ \footnote{Dipartimento di Matematica,
    Universit\`a  di Bologna}}

\date{May 2009}

\maketitle

\begin{abstract}
By using a formal analogy between statistical mechanics of mean
field spin systems and analytical mechanics of viscous liquids -at
first pointed out by Francesco Guerra, then recently developed by
the authors- we give the thermodynamic limit of the free energy
and the critical behavior of Curie Weiss models for a certain
class of generalized spin variables. Then, with the same
techniques, we give a complete picture of the bipartite
Curie-Weiss model, dealing with the same class of generalized
spins. Ultimately we analyze further the existence of a minmax
principle for the latter which mirrors the standard variational
principle of canonical thermodynamics when generalized to multiple
interacting parties.

\begin{center}
\end{center}

\end{abstract}

\section{Introduction}

The investigation of statistical mechanics of mean field spin
systems is experiencing  an increasing interest in the last
decades. The motivations are two-fold: from one side, at the
rigorous mathematical level, a clear picture is still to be
achieved (it is enough to think at the whole community dealing
with the case of random interactions as in glasses \cite{MPV}), at
the applied level, these toy models are starting to be used in
several different context, ranging from quantitative sociology
\cite{CG} to theoretical immunology \cite{parisi}.
\newline
It is then obvious the need for always stronger and simpler
methods to analyze the enormous amount of ''variations on theme'',
the theme being the standard and simplest dichotomic Curie-Weiss
model (CW) \cite{barra0}.
\newline
Recently, inspired by the pioneering work of Francesco Guerra
\cite{sum-rules}, we paved a clear way to manage models with
self-averaging order parameters by using the CW prototype as a
guide \cite{io1}. Here, at first, we apply our scheme to work out
the single-party CW with general spins (i.e. continuous spins with
compact support and symmetric probability measure \cite{GT2})
and we solve in all details its thermodynamics. Then we switch to
the case of  bipartite systems with generalized spins \cite{bip},
both offering a clear picture of the thermodynamics as well as a
digression on the connection of the coupled self-consistent
equations for these models with the existence of an underlying
minmax principle \cite{p.c.}.
\newline
Overall we covered both the ways of investigation: from one side
our analysis is mathematically clear (no powerful but not fully
rigorous methods as replica trick or saddle points are used), from
the other side it is automatically ready for being implemented
into applied scenarios, i.e. the generalized bipartite system
extends competitions in decision making \cite{fadden} among two
communities by allowing their constituent to assume "softer"
viewpoints among each other (with respect to the accept/reject
perspective), being the spins ruling their will continuous instead
of dichotomic.

\section{The ferromagnet with continuously distributed spin variables}
We deal with a system made by $N$ i.i.d. spin r.v. $\sigma_{i}$, $i=1...N$, with probability measure $\mu(\sigma)$, having the following properties:
\newline
\begin{description}

\item[$i)$]$\mu(\sigma)=\mu(-\sigma)$,

\textit{i.e.} it is symmetric;
\newline
\item[$ii)$]$\exists L : \forall \epsilon>0 \int_{L/2}^{L/2+\epsilon}d\mu(\sigma)f(\sigma)=\int_{-L/2-\epsilon}^{-L/2}d\mu{(\sigma)}f(\sigma)=0$,

\textit{i.e.} it has compact support $[-L/2, L/2]$.

\end{description}
In particular, denoting with
$\mathbb{E}_{\sigma_N}=\int_{-\infty}^{+\infty}d\mu(\sigma_1)...d\mu(\sigma_N)$
the expectation values with respect to the $N$ spin variables,  we
notice that from $i)$ it follows that
$\mathbb{E}_{\sigma}[\sigma]=0$, and from $ii)$ that, for a given
bounded function of spin $f(\sigma)$, it has to be
$\mathbb{E}_{\sigma}[f(\sigma)]\leq L[\sup_{\sigma\in[-L/2, L/2]}
f(\sigma)]$.

\bigskip

The spins interact each other, in the way described by the
Hamiltonian $H_N(\sigma,h)$
\begin{equation}\label{eq:hamiltonian}
H_N(\sigma,
h)=-\frac{1}{N}\sum_{(i,j)}^{N,N}\sigma_i\sigma_j-h\sum_i^N\sigma_i.
\end{equation}
Partition function, pressure and free energy per site are defined
as usual as
\begin{eqnarray}
Z_N(\beta, h)&=&\mathbb{E}_{\sigma_N}e^{-\beta H_N(\sigma, h)},\nonumber\\
A_N(\beta, h)&=&\frac{1}{N}\log Z_N(\beta, h),\nonumber\\
f_N(\beta, h)&=&-\frac{1}{\beta}A_N(\beta, h)\nonumber.
\end{eqnarray}
Of course we are interested in calculating the value of the free
energy in thermodynamic limit, \textit{i.e.} for $N\to\infty$, for
describing the thermodynamics of the model. We can also define
Boltzmann states of our system for a generic function of the $N$
spins $g_N$, as
\begin{equation}
\meanv{g_N(\sigma)}=\frac{\mathbb{E}_{\sigma_N}g_N(\sigma)\exp(-\beta H_N(\sigma, h))}{Z_N(\beta, h)}\label{eq:state}.
\end{equation}
It is useful to define also the following quantities:
\begin{eqnarray}
m_N=\frac{1}{N}\sum_i^N\sigma_i,\\ \label{eq:magnet}
a_N=\frac{1}{N}\sum_i^N\sigma_i^2, \label{eq:a_N}
\end{eqnarray}
respectively the magnetization of the system, and the self overlap of spin variables (of course we have trivially $a_N=1$ $\forall N$ in the case of dichotomic spin).
We always have trivially $\meanv{a_N}\leq L^2$.

We can even express the Hamiltonian (\ref{eq:hamiltonian}) in
terms of (3) and (\ref{eq:a_N}). It is
$$
H_N(\sigma, h)=-N(\frac{1}{2}m_N^2+hm_N)+\frac{1}{2}a_N.
$$
This is the starting point of the next section.

\subsection{The free energy in the thermodynamic limit}

We will follow the approach described in
\cite{sum-rules}\cite{io1}. With this purpose, let us introduce
the function $\varphi_N(x,t):
\mathbb{R}\times\mathbb{R^+}\to\mathbb{R}$ defined as
\begin{equation}
\varphi_N(x,t)=-\frac{1}{N}\log\mathbb{E}_{\sigma_N}\exp\left(\frac{tN}{2}m_N^2+xNm_N\right).
\end{equation}

This function plays the role of the pressure (up to a sign)  of a
model defined with a \textit{Boltzmannfaktor} $\exp\left( tN m_N^2
/2+xNm_N \right)$ depending by the two other parameters $(x,t)$
(it is just a matter of names, since both $t$ and $\beta$ as well
 as $x$ and $h$ have the same range of definition). Anyhow it is
strictly related to the pressure of our model, as states the
following

\begin{lemma}
It is, uniformly in $N$
\begin{equation}
|\varphi_N(x,t)+A_N(x,t)|\leq \OO{N}\label{eq:A-phi}.
\end{equation}
Furthermore we have that $\varphi_N(0,x)=A_N(0,x)$ is Lipschitz-continuous $\forall N$.
\end{lemma}

\begin{proof}
Due to convexity of logarithm we get
\begin{eqnarray}
|\varphi_N(x,t)+A_N(x,t)|&=&\frac{1}{N}\left|\log\left[\frac{\mathbb{E}_{\sigma_N}\exp\left(\frac{tN}{2}m_N^2+xNm_N\right)e^{\frac{ta_N}{2}}}{\mathbb{E}_{\sigma_N}\exp\left(\frac{tN}{2}m_N^2+xNm_N\right)}\right]\right|\nonumber\\
&\leq&\frac{1}{N}\log \mathbb{E}_{\sigma}e^{\frac{t a_N}{2}}
\leq\frac{L^2t}{2N}\nonumber,
\end{eqnarray}
and (\ref{eq:A-phi}) is proven. Furthermore, trivially it is
$\varphi_N(0,x)=-A_N(0,x)=-\log\mathbb{E}_{\sigma}[e^{x\sigma}]$.
Again by a convexity argument, joint with the compactness of the
support of the $\sigma$-distribution, it is
\begin{eqnarray}
|\varphi_N(x,0)-\varphi_N(x_0,0)|&=&\left|\log\frac{\mathbb{E}_{\sigma}e^{x\sigma}}{\mathbb{E}_{\sigma}e^{x_0\sigma}}\right|\nonumber\\
&=&\left|\log\frac{\mathbb{E}_{\sigma}e^{(x-x_0)\sigma}e^{x_0\sigma}}{\mathbb{E}_{\sigma}e^{x_0\sigma}}\right|
\leq\log \mathbb{E}_{\sigma}e^{|x-x_0|\sigma} \leq L|x-x_0|,
\end{eqnarray}
hence $\varphi_N(x,0)$ is Lipschitz-continuous. $\Box$
\end{proof}

\begin{remark}
We have that in the thermodynamic limit $A(\beta,
h)=-\varphi(x=h,t=\beta)$.
\end{remark}

\begin{remark}
From (\ref{eq:A-phi}) it is easily seen that in thermodynamic
limit the definitions of state (\ref{eq:state}) and the one built
with Boltzmannfaktor $\exp\left( t N m_N^2 /2+xNm_N \right)$ do
coincide (of course replacing $(\beta, h)$ with $(t,x)$), thus we
actually won't distinguish them in the following.
\end{remark}

The main idea, for solving the thermodynamics encoded into our
Hamiltonian, is to relate the statistical mechanics system to an
effective mechanical one, in which we naturally identify $x$ with
space coordinate, $t$ with time, and the function $\varphi_N(x,t)$
with the mechanical action. In order to do this we notice that
$\varphi_N(x,t)$ satisfies the differential problem
\cite{io1}

\begin{equation}\label{eq:HJ-N-diff}
\left\{
\begin{array}{rclll}
&& \partial_t\varphi_N(x,t)+\frac{1}{2}(\partial_x\varphi_N(x,t))^2-\frac{1}{2N}\partial^2_{x}\varphi_N(x,t)= 0 &\:&\mbox{in }{\mathbb R}\times(0,+\infty)\\
&& \varphi_N(x,0)=-\log\mathbb{E}_{\sigma}e^{x\sigma}&\:&\mbox{on  }{\mathbb
R}\times \{t=0\}.
\end{array}
\right.
\end{equation}

This is a Hamilton-Jacobi equation with a vanishing dissipative
term in the thermodynamic limit. Defined
$u_N(x,t)=\partial_x\varphi_N(x,t)$ the velocity field, we notice
that it corresponds to magnetization of the finite size system in
our parallelism \cite{io1}.We have that $u_N(x,t)$ satisfies a
Burger's equation again with a mollifier dissipative term:

\begin{equation}\label{eq:burgers-N}
\left\{
\begin{array}{rclll}
&& \partial_t u_N(x,t)+ u_N(x,t)\partial_x u_N(x,t)-\frac{1}{2N}\partial^2_{x}u_N(x,t) =
 0&\:&\mbox{in }{\mathbb R}\times(0,+\infty)\\
&& u_N(x,0) =  -\mathbb{E}_{\sigma}\sigma e^{x\sigma}/
\mathbb{E}_{\sigma}e^{x\sigma}&\:&\mbox{on }{\mathbb R}\times
\{t=0\}.
\end{array}
\right.
\end{equation}
Thus the problem of the existence and uniqueness of the
thermodynamic limit is here translated into the convergence of the
viscous mechanical problem to the free one. We can use a theorem,
that resumes a number of results obtained by Peter Lax \cite{lax}
and assures the existence of the solution for free problem:
\begin{theorem}\label{th:lax}
For a general differential problem
\begin{equation}\label{eq:HJ-TL-free}
\left\{
\begin{array}{rclll}
&& \partial_t\varphi(x,t)+\frac{1}{2}(\partial_x\varphi(x,t))^2= 0 &\:&\mbox{in }{\mathbb R}\times(0,+\infty)\\
&& \varphi(x,0)=h(x)&\:&\mbox{on} \ {\mathbb R}\times \{t=0\},
\end{array}
\right.
\end{equation}
and
\begin{equation}\label{eq:burgers-TL-free}
\left\{
\begin{array}{rclll}
&& \partial_t u(x,t)+u(x,t)\partial_x u(x,t)= 0&\:&\mbox{in }{\mathbb R}\times(0,+\infty)\\
&& u(x,0) = g(x)&\:&\mbox{on }{\mathbb R}\times \{t=0\},
\end{array}
\right.
\end{equation}
where $h(x)$ is Lipschitz-continous, and
$g(x)=h^{\prime}(x)\in\mathcal{L}^{\infty}$, it does exist and it
is unique the function $y(x,t):
\mathbb{R}\times\mathbb{R}^+\to\mathbb{R}$ such that
\begin{equation}\label{eq:phi}
\varphi(x,t)=\min_y\left\{
\frac{t}{2}\left(\frac{x-y}{t}\right)^2+h(y) \right\}
=\frac{t}{2}\left(\frac{x-y(x,t)}{t}\right)^2 + h(y(x,t))
\end{equation}
is the unique weak solution of (\ref{eq:HJ-TL-free}), and
\begin{equation}\label{eq:u}
u(x,t)=\frac{x-y(x,t)}{t}
\end{equation}
is the unique weak solution of (\ref{eq:burgers-TL-free}). Furthermore, the function $x\to y(x,t)$ is not-decreasing.
\end{theorem}

It is easily seen that Lax's theorem gives us the solution for the
free energy of the model. In fact, if we put $u(x,t)=-M(x,t)$ in
the solution of the free Burgers' equation, and use $F$ as a short
label standing for ''free'', we get that the minimizing function
is $y(x,t)=x+tM(x,t)$, and the action of the mechanical model
reads off as
$$
\varphi_{F}(x,t)=-A(x,t)=\frac{t^2}{2}M^2(x,t)-\log\mathbb{E}_{\sigma}[\exp\left( \sigma(x+tM(x,t)) \right)].
$$
Therefore, we have the following expression for the free energy per site of our models,
\begin{eqnarray}\label{eq:FE}
f(\beta, h)&=&\frac{1}{\beta}\left[\varphi_F(x,t)\right]_{(t=\beta, x=h)}\nonumber\\
&=&\frac{\beta}{2}M^2(h,\beta)-\frac{1}{\beta}\log\mathbb{E}_{\sigma}[\exp\left( \sigma(h+\beta M(h,\beta)) \right)].\label{eq:FE}
\end{eqnarray}

Finally we must prove convergence of the viscous problem to the
free one. To this purpose, we can state the following
\begin{theorem}
The function
\begin{equation}\label{eq:phi_NbyN}
\varphi_N(x,t)=-\frac{1}{N}\log\sqrt{\frac{N}{t}}\int \frac{dy}{\sqrt{2\pi}}\exp\left[-N\left( (x-y)^2/2t - \log \mathbb{E}_{\sigma}\exp{\sigma y} \right)\right]
\end{equation}
does solve equation (\ref{eq:HJ-N-diff}) and it is
$$
|\varphi_N(x,t)-\varphi_F(x,t)|\leq\OO{N}.
$$
Furthermore, the function
\begin{equation}\label{eq:u_NbyN}
u_N(x,t)=-\frac{\int \frac{dy}{\sqrt{2\pi}}\frac{x-y}{t}\exp\left[-N\left( (x-y)^2/2t - \log \mathbb{E}_{\sigma}\exp{\sigma y} \right)\right]}{\int \frac{dy}{\sqrt{2\pi}}\exp\left[-N\left( (x-y)^2/2t - \log \mathbb{E}_{\sigma}\exp{\sigma y} \right)\right]}
\end{equation}
does solve equation (\ref{eq:burgers-N}) and it is
$$
|u_N(x,t)+M(x,t)|\leq\OO{\sqrt{N}}.
$$
\end{theorem}
The proof is exactly analogue to the one given in \cite{io1}, and
ultimately  due to the uniform convexity of the exponent in
(\ref{eq:phi_NbyN}) and (\ref{eq:u_NbyN}), that we have here by
construction, so we will not report it here.

Therefore we have proven the existence of the thermodynamic limit
for free energy and magnetization of our model.
\newline
As symmetry breaking are fundamental even in statistical
mechanics, we want to report hereafter some other considerations
about the existence and the properties of a phase transition in
our analogy.

\subsection{Phase transition and shock waves}

In this section we deeply study properties of the free Burgers'
equation for the velocity field (\ref{eq:burgers-TL-free}) (that
we remind is the analogue of the magnetization). We can write the
straight line trajectories of the free system (\textit{i.e.} the
system in thermodynamic limit):
\begin{equation}\label{eq:traiettorie-generali}
\left\{
\begin{array}{rrl}
t&=&s\\
x&=&x_0-s \mathbb{E}_{\sigma}\sigma e^{\sigma x_0} /
\mathbb{E}_{\sigma} e^{\sigma x_0}.
\end{array}
\right.
\end{equation}

As usual in these cases, we can find a solution for the
magnetization along characteristics \cite{evans}. It is

\begin{equation}\label{eq:self-cons}
-u(x,t)=M(x,t)=\frac{\mathbb{E}_{\sigma}\sigma\exp\left[ \sigma
(x+tM(x,t))\right]}{\mathbb{E}_{\sigma} \exp\left[ \sigma
(x+tM(x,t))\right]}.
\end{equation}

\begin{remark}
Putting $(x=h, t=\beta)$ in (\ref{eq:self-cons}) we recover the
generalized self consistence equation for the magnetization. In
particular, by choosing $\mu(\sigma) =
(1/2)[\delta(\sigma+1)+\delta(\sigma-1)]$, we immediately
recognize the well known hyperbolic tangent of the dichotomic CW
model.
\end{remark}

An important feature of the velocity field is that it is monotone with respect to $x$. Indeed it is
$$
\partial_x u(x,t)= -\frac{A_{(x,y)}[\sigma^2]}{1+tA_{(x,y)}[\sigma^2]} \leq0,
$$
since $\forall(x,t)$
$$
A_{(x,y)}[\sigma^2]=\frac{\mathbb{E}_{\sigma}\sigma^2\exp\left[ \sigma (x+tM(x,t))\right]}{\mathbb{E}_{\sigma} \exp\left[ \sigma (x+tM(x,t))\right]}-\left(\frac{\mathbb{E}_{\sigma}\sigma\exp\left[ \sigma (x+tM(x,t))\right]}{\mathbb{E}_{\sigma} \exp\left[ \sigma (x+tM(x,t))\right]}\right)^2\geq0
$$

This is known as the entropy condition for the velocity field in
the theory of shock waves \cite{evans}\cite{lax}. For $M(x,t)$ it
follows that
\begin{equation}\label{eq:entropy-cond}
\partial_x M(x,t)\geq0.
\end{equation}

We have seen in \cite{io1} that in usual CW model, \textit{i.e.}
with dichotomic spin variables,  the spontaneous symmetry breaking
associated to the phase transition appears as a shock wave in our
mechanical analogy. The same happens dealing with our generalized
variables.

\begin{proposition}
The line $(t>t_c, 0)$, with $t_c=\sup\frac{M_0}{x_0}$ is a shock
wave for $M(x,t)$, and  by putting
$M_\pm=\lim_{x\to0^\pm}M(x,t)$, it is $M^+=-M^-$.
\end{proposition}

\begin{proof}
With a glance to characteristics (\ref{eq:traiettorie-generali})
we notice that $x=0$ is a stable point of motion\footnote{It is
actually due to the parity of probability measure of $\sigma$.}.
Furthermore for $x=0$ all the straight lines do intersect the
$x$-axis in a certain time. Defining
\begin{equation}\label{eq:t-c}
t_c=\sup_{x_0}\frac{M(x_0, 0)}{x_0}=\sup_{x_0}\partial_xM(x_0,0),
\end{equation}
we have that the line $(t>t_c, 0)$ is a discontinuity line for
$M(x,t)$ since every point on this line is an intersection point
of characteristics, \textit{i.e.} it is a shock waves for the
velocity field $u(x,t)$. We notice from (\ref{eq:t-c}) that, since
we have $\inf (M^2(x,t))=0$, it must be
$$
t_c= \sup_{x_0}\frac{\mathbb{E}_{\sigma}\sigma^2\exp\left[ \sigma (x+tM(x,t))\right]}{\mathbb{E}_{\sigma} \exp\left[ \sigma (x+tM(x,t))\right]} \leq L^2.
$$
On the other hand, we have that for every time there certainly
exists a neighbors of $x=0$ where the function $M(x,t)$ is smooth.
Thus we are allowed to use the Rankine-Hugoniot condition for the
jump along discontinuity \cite{evans}\cite{lax} for stating
$M_+^2=M^2_-$. This last result, coupled with
(\ref{eq:entropy-cond}), completes the proof. $\Box$
\end{proof}

\section{Bipartite models}

We are now interested in considering a set of $N$ spin variables,
in which is precisely defined a partition in two subsets of size
respectively $N_1$ and $N_2$. We assume the variable's label of
the first subset as $\sigma_i$, $i=1, ..., N_1$, while the spins
of the second one are introduced by $\tau_j$, $j=1, ..., N_2$. For
each subset all the spins are i.i.d. r.v., with probability
measure as discussed above, but in principle $\mu(\sigma)$ could
be different by $\mu(\tau)$. Of course we have $N_1+N_2=N$, and we
name the relative size of the two subset $N_2/N_1=\alpha_N$. To
avoid a trivial behavior of the model, we assume that the
thermodynamic limit is performed in such a way that $\alpha=\lim_N
\alpha_N$ is well defined.
\newline
The spins interact via the Hamiltonian $H_N(\sigma, \tau, h_1,
h_2)$:
$$
H_N(\sigma, \tau, h_1,
h_2)=-\frac{1}{N_1}\sum_{i=1}^{N_1}\sum_{j=1}^{N_2}\sigma_i\tau_j
- h_1\sum_{i=1}^{N_1}\sigma_i-h_2\sum_{j=1}^{N_2}\tau_j.
$$
We notice that spins in each subsystem interact only with spins in
the other one, but not among themselves. Partition function,
pressure and free energy per site for the model are defined
naturally, in agreement with the previous section:
\begin{eqnarray}
Z_N(\beta, h_1, h_2)&=&\mathbb{E}_{\sigma_{N_1}}\mathbb{E}_{\tau_{N_2}}e^{-\beta H_N(\sigma, \tau, h_1, h_2)},\nonumber\\
A_N(\beta, h_1, h_2)&=&\frac{1}{N_1}\log Z_N(\beta, h_1, h_2),\nonumber\\
f_N(\beta, h_1, h_2)&=&-\frac{1}{\beta}A_N(\beta, h_1, h_2)\nonumber.
\end{eqnarray}
\begin{remark}
It should be noticed that, for coherence with already known
bipartite models (as i.e. the Hopfield model \cite{BG}), we choose
$N_1$, instead of $N$, as the normalization factor inside the free
energy density and pressure. As we are considering the extensive
scaling among the two parties, i.e. $N_2 = \alpha_N N_1$ and
$\lim_{N \to \infty}\alpha_N = \alpha \in \mathbb{R}^+$, this
simply shifts the overall result by a factor $(1+\alpha)^{-1}$.
\end{remark}
We can also specify the Boltzmann states of our system as
\begin{equation}
\meanv{g_N(\sigma,
\tau)}=\frac{\mathbb{E}_{\sigma_{N_1}}\mathbb{E}_{\tau_{N_2}}g_N(\sigma,
\tau)\exp(-\beta H_N(\sigma, \tau, h_1, h_2))}{Z_N(\beta, h_1,
h_2)}\label{eq:state-bip}.
\end{equation}
As usual, the respective magnetizations of the two systems are
\begin{eqnarray}
m_{N}=\frac{1}{N_1}\sum_i^{N_1}\sigma_i,\\ \label{eq:m-bip}
n_{N}=\frac{1}{N_2}\sum_j^{N_2}\tau_j, \label{eq:n-bip}
\end{eqnarray}
thus the Hamiltonan reads off as
$$
H_N(\sigma, \tau, h)=-N_1\left[\alpha_N m_Nn_N+h_1m_N+h_2\alpha_Nn_N\right].
$$
\subsection{The free energy in the thermodynamic limit}

In order to reproduce the same scheme of the previous section, let us introduce now the $(x,t)$-dependent interpolating partition function
\begin{eqnarray}
&&Z_N(x,t)= \\
&& \nonumber \mathbb{E}_{\sigma}\mathbb{E}_{\tau}\exp N_1 \left(
t\alpha_Nm_Nn_N+\frac{(\beta-t)}{2}(m_N^2+\alpha^2n_N^2)+x(m_N-\alpha_Nn_N)
+h_1m_N+h_2\alpha_Nn_N \right)
\end{eqnarray}
\begin{remark}
Again we notice that the thermodynamical partition function of the
model is recovered when $t=\beta$ and $x=0$.
\end{remark}
We can go further and define the action
\begin{equation}
\varphi_N(x,t)=\frac{1}{N_1}\log Z_N(x,t),
\end{equation}
that therefore is just the pressure of the model for a suitable
choice of $(x,t)$. Now, computing derivatives of $\varphi_N(x,t)$,
we notice that, putting $D_N=m_N-\alpha_Nn_N$, it is
\begin{eqnarray}
\partial_t\varphi_N(x,t)&=&-\frac{1}{2}\meanv{D^2_N}(x,t),\nonumber\\
\partial_x\varphi_N(x,t)&=&\meanv{D_N}(x,t),\nonumber\\
\partial^2_x\varphi_N(x,t)&=&\frac{N_1}{2}\left( \meanv{D_N^2}-\meanv{D_N}^2 \right).\nonumber
\end{eqnarray}
The main difference with respect to the previous case is, instead,
the more complicated form of the boundary condition, \textit{i.e.}
the action at $t=0$. In fact we have that interactions do not
factorize trivially (in a way independent by the size of the
system). It is
\begin{equation}\label{eq:bordo1}
\varphi_N(x,0)=A^1_N(\beta, h_1+x) + \alpha_N A_N^2(\alpha\beta, h_2-x),
\end{equation}
where $A^{1}_N$ is the pressure of the Curie-Weiss model made by
$N_1$ $\sigma$ spins, and $A^{2}_N$ is the same referred to the
$N_2$ $\tau$ spins. Hence, the results of the previous section
give us a perfect control on the function on the r.h.s. of
(\ref{eq:bordo1}), and we have
\begin{equation}\label{eq:bordo2}
\varphi_N(x,0)=A^1(\beta, h_1+x)+\alpha A^2(\alpha\beta, h_2-x)+\OO{N}.
\end{equation}

Thus, again we can build our differential problems for the action $\varphi_N(x,t)$
\begin{equation}\label{eq:HJ-N-bip}
\left\{
\begin{array}{rclll}
&& \partial_t\varphi_N(x,t)+\frac{1}{2}(\partial_x\varphi_N(x,t))^2+\frac{1}{2N_1}\partial^2_{x}\varphi_N(x,t)= 0 &\:&\mbox{in }{\mathbb R}\times(0,+\infty)\\
&& \varphi_N(x,0)=A^1_N(\beta, h_1+x) + \alpha_N A_N^2(\alpha_N\beta, h_2-x)&\:&\mbox{on  }{\mathbb
R}\times \{t=0\},
\end{array}
\right.
\end{equation}
and for the velocity field $D_N(x,t)$
\begin{equation}\label{eq:burgers-N-bip}
\left\{
\begin{array}{rclll}
&& \partial_t D_N(x,t)+D_N(x,t)\partial_x D_N(x,t)+\frac{1}{2N_1}\partial^2_{x}D_N(x,t)= 0&\:&\mbox{in }{\mathbb R}\times(0,+\infty)\\
&& D_N(x,0) = M_N(\beta, h_1+x)-\alpha_N N_N(\alpha_N\beta, h_2-x)&\:&\mbox{on  }{\mathbb R}\times \{t=0\},
\end{array}
\right.
\end{equation}
whence, as for the boundary condition for the action, we have stated
$$
M_N(\beta, h_1+x)-\alpha_N N_N(\beta, h_2-x)=M(\beta, h_1+x)-\alpha N(\alpha_N\beta, h_2-x)+\OO{\sqrt{N}}.
$$
\begin{remark}
We stress that our method, due to the existence of the Burger
equation for the velocity field, introduces by itself the correct
order parameter, without imposing it by hands. We will back on
this point in the last section.
\end{remark}
\begin{remark}
We have that for each collection of values $(\beta, \alpha, h_1,
h_2)$, the function $D_N(x,t)$ is bounded $\forall\:N$,
\textit{i.e.} the function $\varphi_N(x,t)$ is Lipschitz
continuous.
\end{remark}
The main difficulty here is that we have a sequence of
differential problem with boundary conditions dependent by $N$.
Anyway we can replace it with the same sequence of equation but
with fixed boundary condition, that is the well defined limiting
value for $N\to\infty$ of $\varphi_N$ and $D_N$. To this purpose
it is useful the following
\begin{lemma}
The two differential problems
\begin{equation}\label{eq:HJ-N-bip_FIN}
\left\{
\begin{array}{rclll}
&& \partial_t\varphi_N(x,t)+\frac{1}{2}(\partial_x\varphi_N(x,t))^2+\frac{1}{2N_1}\partial^2_{x}\varphi_N(x,t)= 0 &\:&\mbox{in }{\mathbb R}\times(0,+\infty)\\
&& \varphi_N(x,0)=A^1(\beta, h_1+x) + \alpha A^2(\alpha\beta, h_2-x)=h_N(x)&\:&\mbox{on  }{\mathbb
R}\times \{t=0\},
\end{array}
\right.
\end{equation}
and
\begin{equation}\label{eq:HJ-N-bip_FIN}
\left\{
\begin{array}{rclll}
&& \partial_t\bar\varphi_N(x,t)+\frac{1}{2}(\partial_x\bar\varphi_N(x,t))^2+\frac{1}{2N_1}\partial^2_{x}\bar\varphi_N(x,t)= 0 &\:&\mbox{in }{\mathbb R}\times(0,+\infty)\\
&& \varphi_N(x,0)=A^1(\beta, h_1+x) + \alpha A^2(\alpha\beta, h_2-x)=h(x)&\:&\mbox{on  }{\mathbb
R}\times \{t=0\},
\end{array}
\right.
\end{equation}
are completely equivalent, \textit{i.e.} in thermodynamic limit
they have the same solution, $\varphi_N\to\varphi$ and
$\bar\varphi_N\to\varphi$ and it is
$$
|\varphi_N-\bar\varphi_N|\leq\OO{N}.
$$
\end{lemma}
\begin{proof}
By a Cole-Hopf transform, we can easily write the general form of
$\delta_N(x,t)=|\varphi_N(x,t)-\bar\varphi_N(x,t)|$ as
$$
\delta_N=\frac{1}{N}\left|\log\frac{\int_{-\infty}^{+\infty}dy\Delta(y,(x,t))e^{-NR_N(y)}}{\int_{-\infty}^{+\infty}dy\Delta(y,(x,t))}\right|,
$$
where we introduced the modified heat kernel
$\Delta(y,(x,t))=\sqrt{\frac{N}{2\pi t}}\exp\left(-N\left[
(x-y)^2/2t+h(y)\right]\right)$, and $R_N(y)=|h(y)-h_N(y)|$, with
$\lim_N NR_N<\infty$, $\forall\:y$. Now we notice that because of
theorem 2, it certainly exists an $y^*$ such that
$$
\sup_y R_N(y)=y^*\qquad\mbox{ and }\qquad\lim_N NR_N(y^*)<\infty.
$$
Hence it is
\begin{eqnarray}
\delta_N(x,t)&\leq& \frac{1}{N} |\log e^{-NR_N(y^*)}|\nonumber\\
&=&\frac{1}{N}\left[NR_N(y^*)\right]\leq\OO{N},
\end{eqnarray}
that completes the proof.$\Box$
\end{proof}

Of course a similar result holds also for the Burgers' equation for the velocity field $D_N$.

So, finally, we must study
\begin{equation}\label{eq:HJ-N-bip_FIN}
\left\{
\begin{array}{rclll}
&& \partial_t\varphi_N(x,t)+\frac{1}{2}(\partial_x\varphi_N(x,t))^2+\frac{1}{2N_1}\partial^2_{x}\varphi_N(x,t)= 0 &\:&\mbox{in }{\mathbb R}\times(0,+\infty)\\
&& \varphi_N(x,0)=A^1(\beta, h_1+x) + \alpha A^2(\alpha\beta, h_2-x)&\:&\mbox{on  }{\mathbb
R}\times \{t=0\},
\end{array}
\right.
\end{equation}
and
\begin{equation}\label{eq:burgers-N-bip_FIN}
\left\{
\begin{array}{rclll}
&& \partial_t D_N(x,t)+D_N(x,t)\partial_x D_N(x,t)+\frac{1}{2N_1}\partial^2_{x}D_N(x,t)= 0&\:&\mbox{in }{\mathbb R}\times(0,+\infty)\\
&& D_N(x,0) = M(\beta, h_1+x)-\alpha N(\alpha\beta, h_2-x)&\:&\mbox{on  }{\mathbb R}\times \{t=0\}.
\end{array}
\right.
\end{equation}
Now the path is clear, and we can state the following

\begin{theorem}
The pressure of the generalized bipartite ferromagnet, in
 the thermodynamic limit, is given by:
\begin{equation}\label{eq:A_BIP}
A(\beta, \alpha, h_1, h_2)=-\alpha\beta\tilde N\tilde M + \log\mathbb{E}_{\sigma}\exp\left[\sigma\left(h_1+\alpha\beta \tilde N\right)\right]+\alpha\log\mathbb{E}_{\tau}\exp\left[\tau\left(h_2+\beta \tilde M\right)\right],
\end{equation}
where, given the well defined magnetization for the generalized CW
model respectively for $\sigma$ and $\tau$, $M(\beta, h)$ and
$N(\beta, h)$, it is
\begin{eqnarray}
\tilde M(\beta, \alpha, h_1, h_2)&=&M(\beta, h_1-\beta M+\alpha\beta N)\label{eq:Mtilde}\\
\tilde N(\beta, \alpha, h_1, h_2)&=&N(\beta, h_2+\beta M-\alpha\beta N)\label{eq:Ntilde}.
\end{eqnarray}
Furthermore it is
\begin{equation}\label{eq:conv_bip}
|A_N(\beta, h_1, h_2)-A(\beta, \alpha, h_1, h_2)|\leq\OO{N}.
\end{equation}
\end{theorem}
\begin{proof}
Theorem 1 gives us the existence and the form of the free
solution. We know \cite{evans} that the free Burger's equation can
be solved along the characteristics
\begin{equation}\label{eq:traiettorie-generali_BIP}
\left\{
\begin{array}{rrl}
t&=&s\\
x&=&x_0+sD(x_0, 0),
\end{array}
\right.
\end{equation}
where
$$
D(x_0, 0)=M(\beta, h_1+x_0)+\alpha N(\alpha\beta, h_2-x_0),
$$
and it is
$$
D(x,t)=D(x_0(x,t),0)=M(\beta, h_1+x-tD(x_0,0))+\alpha N(\alpha\beta, h_2-x+tD(x_0,0)).
$$
Then we can notice that
\begin{eqnarray}
M(\beta, h_1+x-tD(x_0,0))&=&\frac{\mathbb{E}_{\sigma}\sigma\exp\left[\sigma\left(h_1+x+t\alpha N\right)\right]}{\mathbb{E}_{\sigma}\exp\left[\sigma\left(h_1+x+t\alpha N\right)\right]}\label{eq:self-M-bip},\\
N(\alpha\beta, h_2-x+tD(x_0,0))&=&\frac{\mathbb{E}_{\tau}\tau\exp\left[\tau\left(h_2-x+t M\right)\right]}{\mathbb{E}_{\tau}\exp\left[\tau\left(h_2-x+t M\right)\right]}\label{eq:self-N-bip},
\end{eqnarray}
which coincide with (\ref{eq:Mtilde}) and (\ref{eq:Ntilde}) when
$x=0$ and $t=\beta$.

At this point we know that the minimum in theorem 1 is taken for
$y=x-tD(x,t)$, and, bearing in mind the general form of the
pressure of CW models, given in the last section, we have
\begin{eqnarray}
\left[\varphi(x,t)\right]_{(x=0,t=\beta)}&=&\Big[\frac{t}{2}D^2(x,t)-\frac{t}{2}M^2(\beta, h_1+x-tD(x_0,0))-\frac{t}{2}\alpha^2 N^2(\alpha\beta, h_2-x+tD(x_0,0))\nonumber\\
&+& \log\mathbb{E}_{\sigma}\exp\left[\sigma\left(h_1+x+t\alpha N\right)\right]+\alpha\log\mathbb{E}_{\tau}\exp\left[\tau\left(h_2-x+t M\right)\right]\Big]_{(x=0,t=\beta)}\nonumber\\
&=&A(\beta, \alpha, h_1, h_2),\nonumber
\end{eqnarray}
where $A(\beta, \alpha, h_1, h_2)$ is given just by
(\ref{eq:A_BIP}), bearing in mind the right definition of $\tilde
M$ and $\tilde N$. Now we must only prove the convergence of the
true solution to the free one. But, exactly like in theorem 2,
equation (\ref{eq:conv_bip}) follows by standard techniques,
because of the uniform concavity of
$$
\frac{(x-y)^2}{2t}+A^1(\beta, h_1+y)+\alpha A^2(\beta, h_2-y)
$$
with respect to $y$, assured by theorem 1. In fact we have that,
by a Cole-Hopf transform \cite{evans}, the unique bounded solution
of the viscous problem is
$$
\varphi_N(x,t)=\frac{1}{N}\log\sqrt{\frac{N}{t}}\int \frac{dy}{\sqrt{2\pi}}\exp\left[-N\left( \frac{(x-y)^2}{2t}+A^1(\beta, h_1+y)+\alpha A^2(\beta, h_2-y) \right)\right]
$$
and we have, by standard estimates of a Gaussian integral, that
$$
\left|\varphi(x,t)-\varphi_N(x,t)\right|\leq\OO{N},
$$
\textit{i.e.} also the (\ref{eq:conv_bip}) is proven.$\Box$
\end{proof}
Finally, by this last theorem, we can easily write down the free energy of the model:
$$
f(\alpha, \beta, h_1, h_2)=\alpha\tilde N\tilde M - \frac{1}{\beta}\log\mathbb{E}_{\sigma}\exp\left[\sigma\left(h_1+\alpha\beta \tilde N\right)\right]-\frac{\alpha}{\beta}\log\mathbb{E}_{\tau}\exp\left[\tau\left(h_2+\beta \tilde M\right)\right].
$$
\begin{remark}
We stress that when recovering the one party scenario (i.e.
$\alpha=0$) the model trivially reduces to the well known CW in an
external magnetic field, with the free energy $-\beta
f(\beta,h_1)= \ln 2 + \ln\cosh(\beta h_1)$.
\end{remark}
In the last paragraph we will see how expressions like this one
can be derived thought a minmax principle.

\subsection{The occurrence of a minmax principle for the free energy}

As we have seen in the previous paragraph, the velocity field
$D_N(x,t)$ plays the role of order parameter for the model.
Actually, in perfect analogy with other cases of interest (see for
instance the last section about generalized ferromagnets, or
\cite{io1}), the free energy is then obtained minimizing (or
maximizing, depending on the complexity of the system, i.e. the
presence of frustration \cite{MPV}) the action with respect to the
order parameter. In bipartite model one has two natural order
parameters, \textit{i.e.} each of which referred to the party it
belongs to. From our study of bipartite ferromagnet, we know that
the true order parameter is a linear combination of the two
magnetizations, one for each parties, $D=M-\alpha N$: What is done
by Lax's theorem, for example for the free energy, is taking the
maximum of $D$ on a suitable trial functional \cite{p.c.}
\begin{eqnarray}
f(\alpha, \beta, h_1, h_2)&=&\max_{D}\Big[ -\frac{D^2}{2}+\frac{M^2}{2}+\frac{\alpha^2N^2}{2}\nonumber\\
&-&\frac{1}{\beta}\log\mathbb{E}_{\sigma}\exp\left[\sigma\left( h_1+\beta M +\beta D \right)\right] - \frac{\alpha}{\beta}\log\mathbb{E}_{\tau}\exp\left[\tau\left( h_2+\alpha\beta N -\beta D \right)\right] \Big]\nonumber
\end{eqnarray}
This expression is rather unsatisfactory, since not only the order
parameter of the model $D$ appears, but even the two
magnetizations $M$ and $N$. Anyway we can see the model as
described by two different order parameters, $M$ and $N$
themselves, and in the last expression one should take the
extremum with respect to both $M$ and $N$. Anyway we have that
$D=M-\alpha N$, thus maximize $D$ is equivalent to maximize $M$
and  minimize $N$. We must only rewrite our trial functional in
terms of $M$ and $N$, and we have the minmax principle for the
free energy
$$
f=\min_{N}\max_{M}\Big[ \alpha MN -\frac{1}{\beta}\log\mathbb{E}_{\sigma}\exp\left[\sigma\left( h_1+\alpha\beta N \right)\right] - \frac{\alpha}{\beta}\log\mathbb{E}_{\tau}\exp\left[\tau\left( h_2+\beta M  \right)\right] \Big].
$$
It naturally arises from the last formula that the free energy is concave with respect to $N$ and convex with respect to $M$, but of course it is uniformly convex along $M-\alpha N$. Indeed, we have that $M$ and $N$ are not independent, but are related by (\ref{eq:Mtilde}) and (\ref{eq:Ntilde}), that is
\begin{equation}
M=\frac{\mathbb{E}_{\sigma}\sigma\exp\left[\sigma\left(h_1+\alpha\beta
N\right)\right]}{\mathbb{E}_{\sigma}\exp\left[\sigma\left(h_1+\alpha\beta
N\right)\right]}, \  \  \
N=\frac{\mathbb{E}_{\tau}\tau\exp\left[\tau\left(h_2+\beta
M\right)\right]}{\mathbb{E}_{\tau}\exp\left[\tau\left(h_2+\beta
M\right)\right]}.
\end{equation}

\begin{remark}
As for the single party model, we stress that when choosing
$\mu(\sigma)= (1/2)[\delta(\sigma+1)+\delta(\sigma-1)]$, i.e.
dichotomic case, the self-consistent relations reduce to the
already known\cite{bip}
\begin{eqnarray}
M(\beta, h_1, \alpha, N) &=& \tanh\large( h_1 + \beta\alpha N \large), \\
N(\beta, h_1, \alpha, N) &=& \tanh\large( h_2 + \beta M \large).
\end{eqnarray}
However, with respect the model analyzed in \cite{bip} it should
be noticed that we miss the self-contribute inside each equation
(i.e. $M \neq f(M)$ as well as $N \neq f(N)$). This is ultimately
due to the lacking of the self-interaction inside each party into
the Hamiltonian we are considering.
\end{remark}

These are the true self-consistence relations of the model,
analogue to (\ref{eq:self-cons}), and we conclude that the choice
of two different order parameters is redundant, since they are
related. One might make the choice of putting $N=N(M)$ and study
the problem using only $M$ as order parameter (or viceversa), but,
as we have seen it is not so convenient, since a beautiful
extremum principle does not seem to arise studying the system
along the direction of one of the two subsystems\footnote{This is
finally due to the symmetry between the $\sigma$ subsystem and the
$\tau$ one.}, \textit{i.e.} along $M$ or $N$. In fact we know,
thanks to our technique, that the extremum is taken with respect
to $D$.

Thus actually one has only one degree of freedom, and the minmax
principle, although on one hand it gives a more satisfactory form
of the flow equations, on the other hand only hides a more
meaningful minimum or maximum principle. This characteristic of
bipartite model seems to be quite general, and might be extended
to other models of interest in future development.

\section{Conclusion}

In this paper we used a mechanical analogy, introduced and
developed in \cite{sum-rules}\cite{io1}, for a complete resolution
of mean field ferromagnetic models with a very general class of
spin r.v., \textit{i.e.} with probability measure symmetric and
with compact support. The free energy in the thermodynamic limit
and the phase transition have appeared in our work as,
respectively, the solution in the limit of vanishing viscosity of
a Hamilton-Jacobi equation with diffusion, and the occurrence of a
shock line for the related velocity field. Moreover, we have
applied the same methods to the more interesting bipartite
systems, made by two different subsystem of spins (a priori of
different nature), each one interacting with the other, but with
no self-interactions. We have seen that the thermodynamic limit of
the pressure does exist and it is unique and we gave its explicit
expression in a constructive way. Further, when introducing the
Burger's equation for the velocity field, our methods
automatically ''choices'' the proper order parameter, which turns
out to be a linear combination of the magnetizations of the two
subsystems, with different signs. By this property, we developed
an analysis of the minmax principle, pointing out its importance
relating it to the more classical min/max for the free energy (or,
of course, for the pressure) for this very simple model. Noticing
that the same structure can be recovered for many other models of
greater interest, like bipartite spin glasses,  we plan to  report
soon about them.

\section*{Acknowledgements}

Authors are grateful to Francesco Guerra for his priceless
scientific guide. Moreover they would like to thank DSNN group of
King's College London and Renato Luc\`a for useful discussions.
\newline
Authors furthermore are grateful to EDENET Onlus for a grant
permitting this work.
\newline
AB travels are covered via a GNFM grant which is also
acknowledged.



\end{document}